\documentclass[aps,pre,twocolumn,groupedaddress,showpacs,floatfix]{revtex4}
\usepackage{amsmath}
\usepackage{amssymb}
\usepackage{graphicx}

\begin{document}

\title{On the Dynamics of the $h-$index in Complex Networks
with Coexisting Communities}

\author{Luciano da Fontoura Costa}
\affiliation{Instituto de F\'{\i}sica de S\~ao Carlos.
Universidade de S\~ ao Paulo, S\~{a}o Carlos, SP, PO Box 369,
13560-970, phone +55 16 3373 9858,FAX +55 16 3371 3616, Brazil,
luciano@if.sc.usp.br}

\date{9th Sep 2006}

\begin{abstract}
This article investigates the evolution of the $h-$index in a complex
network including two communities (in the sense of having different
features) with the same number of authors whose yearly productions
follow the Zipf's law.  Models considering indiscriminate citations,
as well as citations preferential to the fitness values of each
community and/or the number of existing citations are proposed and
numerically simulated.  The $h-$indices of each type of author is
estimated along a period of 20 years, while the number of authors
remains constant.  Interesting results are obtained including the fact
that, for the model where citations are preferential to both community
fitness and number of existing citations per article, the $h-$indices
of the community with the largest fitness value are only moderately
increased while the indices of the other community are severely and
irreversibly limited to low values.  Three possible strategies are
discussed in order to change this situation.  In addition, based on
such findings, a new version of the $h-$index is proposed involving
the automated identification of virtual citations which can provide
complementary and unbiased quantification of the relevance of
scientific works.
\end{abstract}

\maketitle

\emph{`The only factor becoming scarce in a world of abundance
is human attention.'}  (Kevin Kelly, Wired) \vspace{0.5cm}

\section{Introduction}

It all started in darkness and mystery.  In the beginnings of
humankind, explanation of the world and prediction of the future
lied deep into the impenetrable realm of sorcerers and medicine men.
Except for the extremely rare initiated, the inner ambiguous
workings of divination and sorcery were jealously guarded.  Similar
secrecy was observed through much of the subsequent history of
humankind, including the age of oracles in the classical world and
alchemy all through the middle ages.  `Knowledge' was not for
everybody, it was the power source of a few.  Ultimately, the value
of those practices did not stem from their effectiveness, but
emanated from all types of symbology, dogmas, metaphors and
ambiguities.

With time, some light was shed, and part of humankind finally realized
the value of confronting explanations and predictions with reality,
through experimentations. That such a basic fact would take such a
long time to be inferred provides a lasting indication of the inherent
limitations of human nature.  Be that as it may, the value of
experiments finally established itself, from the renaissance up to the
present day.  Such an essential change was accompanied by another
important fact: it became progressively clearer that once widely
disseminated, new findings acted in order to catalyze still more
discoveries. The popularization of printing techniques contributed
substantially to implementing this new philosophy, being steadily
crystalized into an ever growing number of books, and then journals
and WWW files.  One of the immediate consequences of the first
scientific papers was the respectively unavoidable \emph{citations}.
Today, citations and impact factors (calculated by taking into account
citations, as well as other indicators) are widely used, to the
happiness of some and chagrin of others, for quantifying the quality
and productivity of researchers, groups, institutions and journals.

Scientific indices~\cite{Garfield:72} are now regularly applied in
order to decide on promotions, grants and identification of scientific
trends.  In this way, science became, to a large extent, driven by
scientometry.  However, it is important not to forget the initial
purpose of scientific publishing of \emph{fostering dissemination of
high quality knowledge and results for the benefit of humankind}.  One
important point to be born in mind refers to the fact that all
existing scientific indices are biased in some specific way. For
instance, the total number of articles published by a researcher is
not necessarily related to its productivity unless their age (or
seniority) is taken into account. At the same time, the number of
citations received by a work or author is also relative, because this
number can depend on joint-authorship, the specific area, or even be a
consequence of some error in the original work.  Yet, though not
perfect, scientific indices do provide some quantification of the
productivity and quality of papers, researchers, institutions and even
countries and continents.  The common sense approach, given the
unavoidable limitations of the indices, is not to dismiss them, but to
try to identify their faults so that they can be further improved.
And, little wonder, the best bet is to use \emph{science} to improve
the \emph{scientific} indices.

It is a positive sign of our age that relatively great effort,
reflected in a growing number of related publications
(e.g.~\cite{Garfield:72, nature_news:2005, Batista_etal:2007,
Bornmann:2005, Sidiropoulos:2006, Egghe_dynamic:2006,
Egghe_improvement:2006, Borner:2004, Hirsch:2005, Popov:2005,
Miller:2006, Cronin_Meho:2006,Raan:2005}), that science has indeed
been systematically used for studying and enhancing scientific
indices.  One of the most interesting recent developments was Hirsh's
proposal of the $h-$index~\cite{Hirsch:2005}. Having sorted the
articles of a researcher in decreasing order of citations, the $h$
value can be immediately obtained as the position $h$ along the
sequence of articles just before the number of citations become
smaller than $h$. Several are the advantages of such an
index~\cite{Hirsch:2005, nature_news:2005,Cronin_Meho:2006} with
respect to other more traditional indicators, including the fact that
the $h-$index does not take into account the citation tail (i.e. works
with few citations) and is more robust than the total number of
citations per author in the form of sporadic joint publications with
famous researchers~\footnote{Though an author may have harvested as
many as 1000 citations from a single jointly written article, this
entry alone will imply an $h-$index equal to 1}.  However, the
$h-$index has also been acknowledged by Hirsh to be potentially biased
by factors such as the number of authors and the specific scientific
area~\cite{Hirsch:2005}.  Several additional specific shortcomings of
the $h-$index have been identified and efforts have been made at their
respective correction (e.g.~\cite{Egghe_improvement:2006,
Bornmann:2005, Batista_etal:2007, Sidiropoulos:2006}).  Yet, the
$h-$index is indeed an interesting alternative which deserves
continuing attention aimed at possible further refinements.  Growing
attention has also been focused on the dynamical aspects of the
evolution of the $h=$index (e.g.~\cite{Egghe_dynamic:2006}) as well as
the joint consideration of the evolution of author and articles
networks (e.g.~\cite{Borner:2004}).  Another interesting trend is the
comparison of the $h-$index with more standard scientometrics
indicators including peer judgements (e.g.~\cite{Bornmann:2005,
Raan:2005}).

The total number of publications of an author can be roughly estimated
from the $h-$index as $C_T = a h^2$~\cite{Hirsch:2005}, where $a$ is a
constant empirically found to lie between $3$ and $5$. In other words,
though not precise, this relationship explains a good deal about the
source of the greater stability of this measurement when compared to
the traditional total number of citations $C_T$. At the same time, the
above relationship is not perfect, otherwise the $h-$index would be
but a transformed version of the total number of citations.  Another
interesting measurement proposed by Hirsch~\cite{Hirsch:2005} is the
$m-$index, defined by the linear model $h \propto mn$, where $n$ is
the number of subsequent ages (usually years).  Therefore, $m$
corresponds to the approximate (mean or instant) rate of increase of
the $h-$index with time.  An $m-$index of 3 obtained for a researcher,
for instance, suggests that his $h-$index tends to increase 30 times
after 10 years.

Several related investigations, including Hirsh's original
work~\cite{Hirsch:2005}, assume that the articles tend to receive a
fixed number of citations $c$ along time.  While it would be possible
to consider a time window for citations, it is also interesting to
take into account \emph{preferential} citation rules such as in
complex networks research (e.g.~\cite{Albert_Barab:2002,
Boccaletti:2006, Newman:2003, Costa_surv:2006}).  According to this
model, nodes which have many connections (e.g. citations) tend to
attract more connections, giving rise to the `rich get richer'
paradigm. Another important aspect which has been relatively
overlooked is the presence of communities in the scientific world
(e.g.~\cite{Newman:2003, Costa_surv:2006}). Several are the possible
origin for such communities, including the area of research, language
of publication, age, style, among many others.

The present work reports an investigation on the simulated dynamics of
the $h-$index considering variable number $c$ of citations received
per article, defined by preferential attachment.  As such, this work
represents one of the first approaches integrating $h-$index and
complex networks.  However, we believe its main contributions to lie
elsewhere, mainly in the consideration of the two
communities~\footnote{It should be observed that the term community is
used in this work in order to identify two subsets of nodes which
share some features (i.e. fitness), rather than in the sense of being
more interconnected one another than with the remainder of the
network.}, henceforth abbreviated as A and B, with distinct fitness
values and under the realistic dynamics of preferential attachment, as
well as the assumption that the number of papers published by each
author follows the Zipf's law (e.g.~\cite{Newman_Zipf:2004}).  These
two communities produce articles with respective fixed fitness indices
$f_A$ and $f_B$.  In order to reflect some inherent difference between
the two communities -- e.g. as a consequence of the researcher age,
writing style, language or specific area (more likely combinations of
these) -- we impose that $f_A=2f_B$, i.e. the articles in community A
are inherently twice as much more citable than those produced by the
other community. Note that any of the above criterion can be used to
separate the citation networks into 2 or more subgraphs, e.g. by
establishing respective thresholds~\footnote{For instance, if age is
to be considered as a parameter, community A could be obtained by
selecting those nodes (and respective edges) corresponding to authors
older than $T$ years, with the remainder nodes defining community B.}.
It is also important to emphasize at the outset that the presence of
these two (or more) communities is \emph{assumed} rather than taken
for granted.  The same can be said about the possible origin of the
fitness difference.  It is hoped that the present work can provide
subsidies for the eventual identification of such distinct communities
from the perspective of the observation of the $h-$indices of the
respective authors.

The considered simulated dynamics extends over 20 years.  Because of
computational restrictions, the number of authors is limited to 78
which, under the Zipf's law, implies a total of 302 papers per year.
Each article is assumed to yield the fixed number of $w$ citations to
other works, self-citations included. For simplicity's sake, the
number or papers published per year by each author, as well as the
number of authors, are also considered fixed, which is not a great
drawback given the relatively short period of the simulation (i.e.  20
years). Despite its unavoidable simplifications, the suggested model
provides a number of remarkable results and trends, including bleak
perspectives for the community with smaller fitness (B), which are
identified and discussed.  A brief discussion is also provided
concerning possible strategies to be adopted by community B in order
to improve its overall $h-$indices.  Based on the simulation results,
the proposal for yet another enhanced version of the $h-$index, based
on the identification of virtual citations in terms of the number of
shared main features of each work (e.g. revelabed by statistics or
artificial intelligence), is outlined.

The article starts by defining the model and follows by presenting the
results obtained considering two values of $w$ and
uniform/preferential attachment.  Considerations are made regarding
possible means to change the situation of community B as well as for
the proposal of a new version of the $h-$index.  The work concludes by
summarizing the main findings and suggesting perspectives for future
works.

\section{The Models}

The number of articles published by year by each author $i$,
henceforth $y(i)$, is assumed to follow the Zipf's
distribution~\cite{Newman_Zipf:2004}, i.e.

\begin{equation}  \label{eq:py}
  p(y) = c y^\beta,
\end{equation}

where $p(y)$ is the distribution probability of $y$ and $\beta$ and
$c$ are real parameters.  We define the specific form of this relation
(i.e. its parameters) by establishing the two extremity points
$(y,p(y))$ and $(1,m)$ and $(s,1)$ of the respective distribution.  In
other words, we assume that $m$ authors publish only one paper per
year and only one author publishes $s$ papers per year.  Therefore, we
have that

\begin{eqnarray}
  c = m, \\
  \beta = -log(m)/log(s)
\end{eqnarray}

It is henceforth assumed that $m=15$ and $s=30$.  In addition, we
have to sample from this distribution.  Without great loss of
generality, we chose $y = (1, 2, 3, 5, 10, 15, 30)$ and consequently
obtain $p(y) =(15, 9, 6, 4, 2, 2, 1)$.  In other words, 15 authors
publish one article per year, 9 authors publish 2 articles per year,
and so on. This leads to $NA=39$ authors and a total of $151$ yearly
articles. Such a configuration is assumed for the two considered
communities A and B, implying a grand total of 78 authors and 302
papers per year.

In order to represent the citations network, we adopt a directed
network (i.e. a digraph) defined as $\Gamma=(V,Q)$, where $V$ is the
set of $N$ vertices (or nodes) representing the articles and $Q$ is
the set of $E$ edges (or links) connecting the nodes (i.e. the
citations). Note that both $V$ and $Q$ vary along the 20
considered years.  A citation from an article $j$ to another article
$i$ is represented as $(j,i)$ and stored into the adjacency matrix
$K$ as $K(i,j)=1$ (a null entry is imposed otherwise). The number of
citations received by each article $i$ is immediately given in terms
of the respective \emph{indegree} of the respective node, i.e.

\begin{equation}
  k(i)=\sum_{r=1}^{N} K(i,r)
\end{equation}

Although presenting identical structure as far as the number of
authors and respective number of articles published per year are
concerned, the fitness of the articles produced by community A can be
considere to be twice as large as those published by community B,
i.e. $f_A=2f_B$. We henceforth assume that$f_B=1$.  These values are
used in order to bias the establishment of the links during the
simulations as explained below.

The growth of the citation network is performed in yearly terms.
Four dynamics are considered for comparison purposes: (i) \emph{UNI
-- uniform}; (ii) \emph{PREFF -- preferential to community fitness};
(iii) \emph{PREFC -- preferential to existing article citations};
and (iv) \emph{DBPREF -- preferential to community fitness and
existing citations}, each of which is described in the following.
Though all models considered in this article do not include a
citation time window, this is not a great shortcoming given the
relatively short period of the simulation (i.e. 20 years).

In the \emph{UNI} model, each of the 301 articles added each year are
assumed to cite exactly $w$ articles randomly chosen among those
published from the previous to the current year. We consider two
situations, defined by $w=5$ and $20$.  The \emph{PREFF} model is
similar to the UNI scheme, but now the new citations take into account
the communities fitness. As a consequence, articles from community A
become twice as much likely to be cited than those from community
B. The \emph{PREFC} model is also preferential, but here each of the
$w$ citations per article is performed preferentially to the number of
existing citations of each article published from the beginning to the
current year.  This model is therefore similar to the
Barab\'asi-Albert model (e.g.~\cite{Albert_Barab:2002, Newman:2003,
Boccaletti:2006, Costa_surv:2006}), except that the indegrees
(i.e. number of citations) are not updated during the year, but only
at its end. Finally, the \emph{DBPREF} model is doubly preferential,
to both existing citations and communities. More specifically, a list
is kept where the identification of each article is entered a total
number of times corresponding to the value of its incoming citations
multiplied by the community fitness (i.e. $f_A=2$ for community A and
$f_B=1$ for community B). New citations are then chosen by random
uniform selection among the elements in the above list.  Each of the
configurations was run 50 times in order to provide statistical
representativeness, while the $h-$index and total number of citations
per author $N_T$ were calculated for each author at each year.

\section{Simulation Results and Discussion}

\begin{figure*}[h]
 \begin{center}
   \includegraphics[scale=0.5,angle=0]{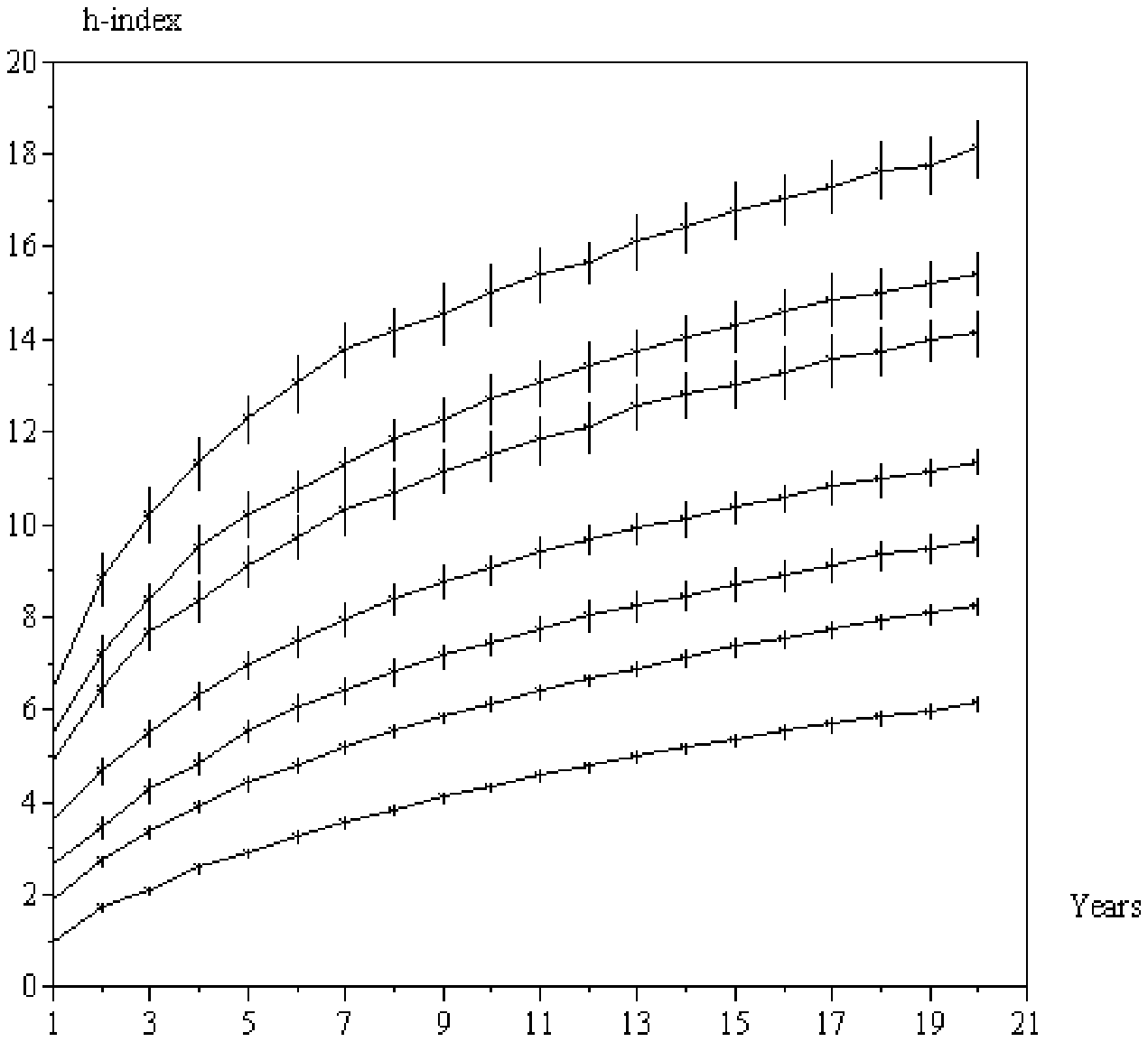} \hspace{1cm}
   \includegraphics[scale=0.5,angle=0]{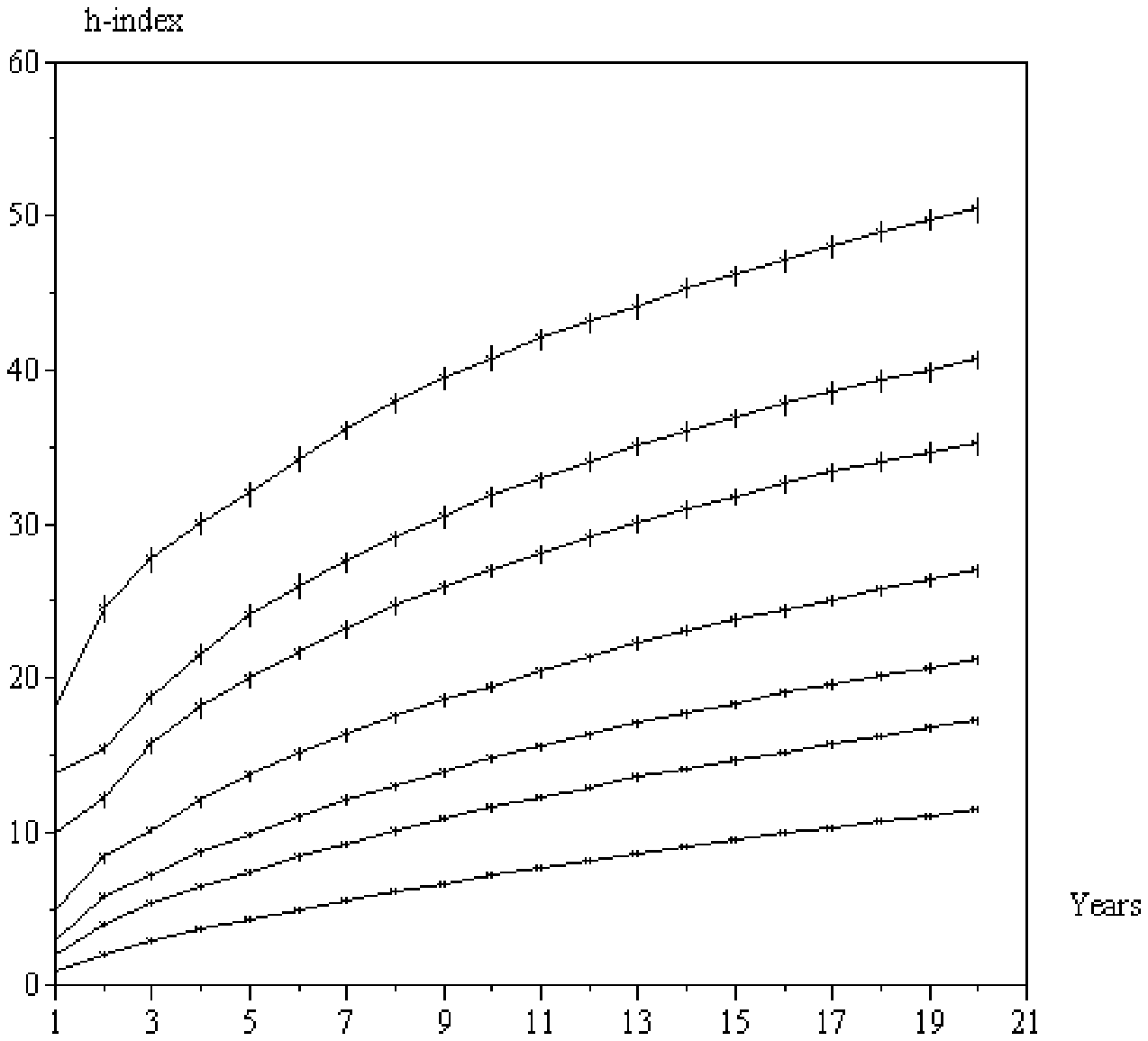} \\
   (a)  \hspace{8cm}   (b)  \\
   \includegraphics[scale=0.5,angle=0]{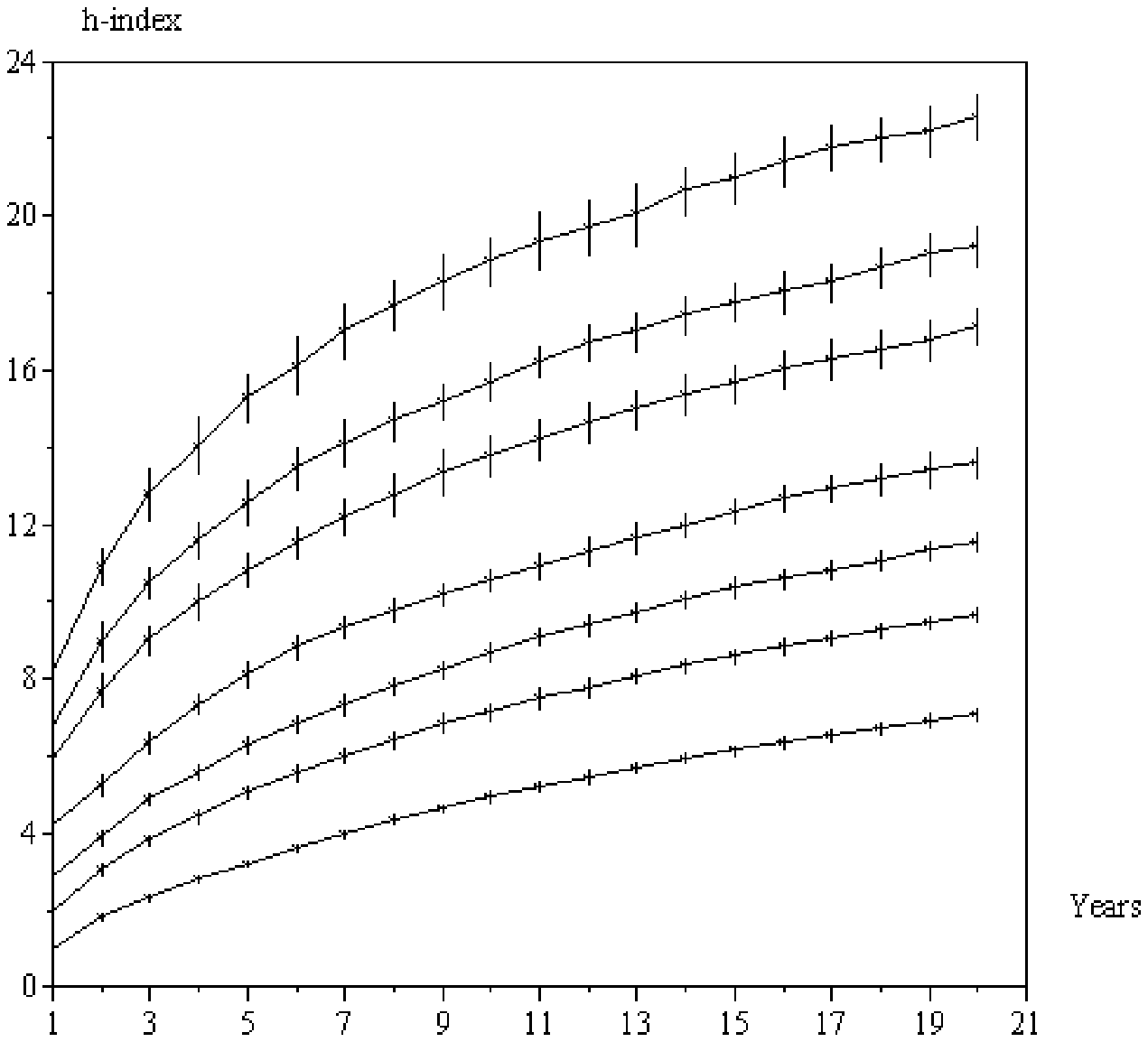}  \hspace{1cm}
   \includegraphics[scale=0.5,angle=0]{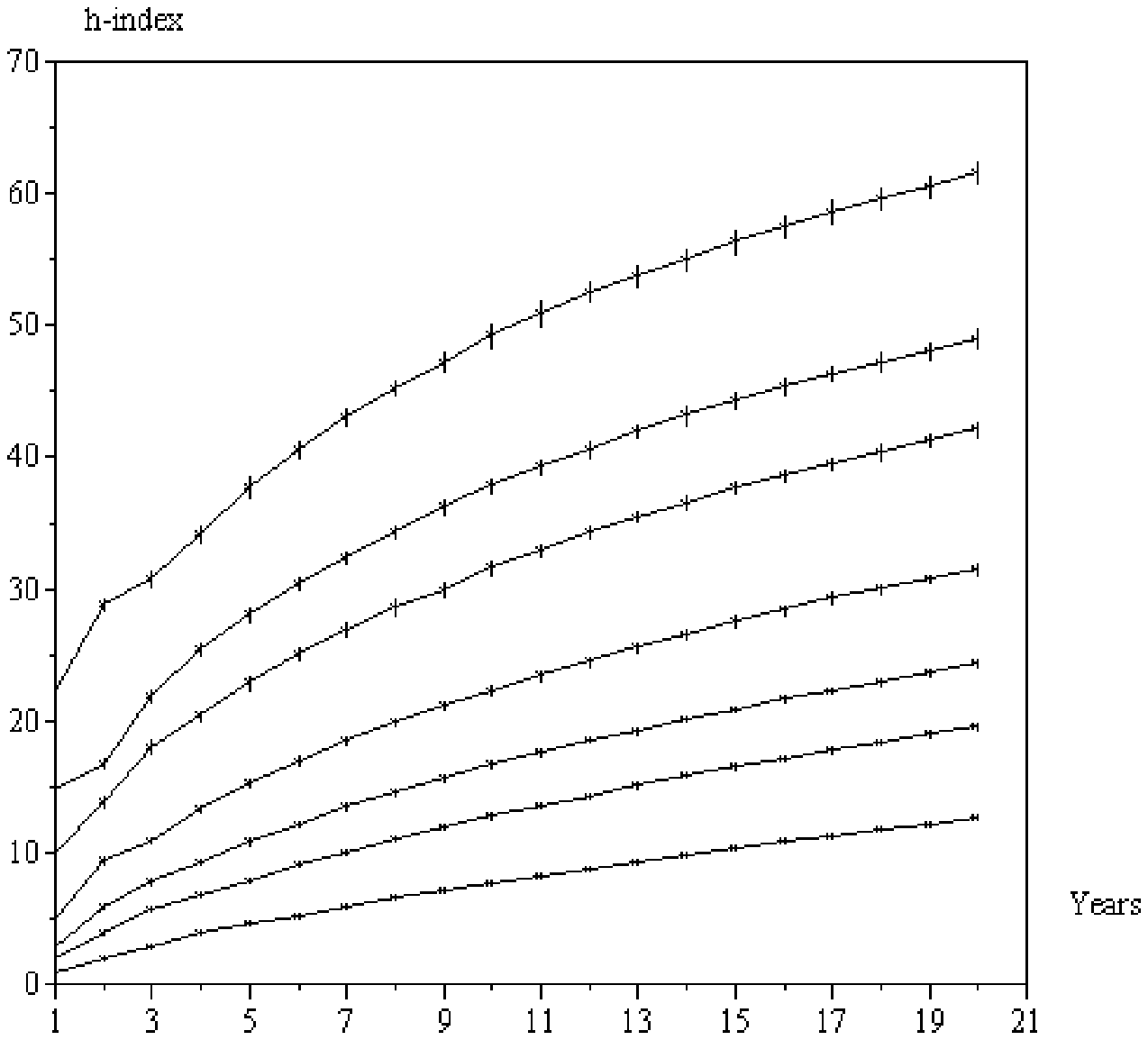} \\
   (c)   \hspace{8cm}   (d)  \\
   \includegraphics[scale=0.5,angle=0]{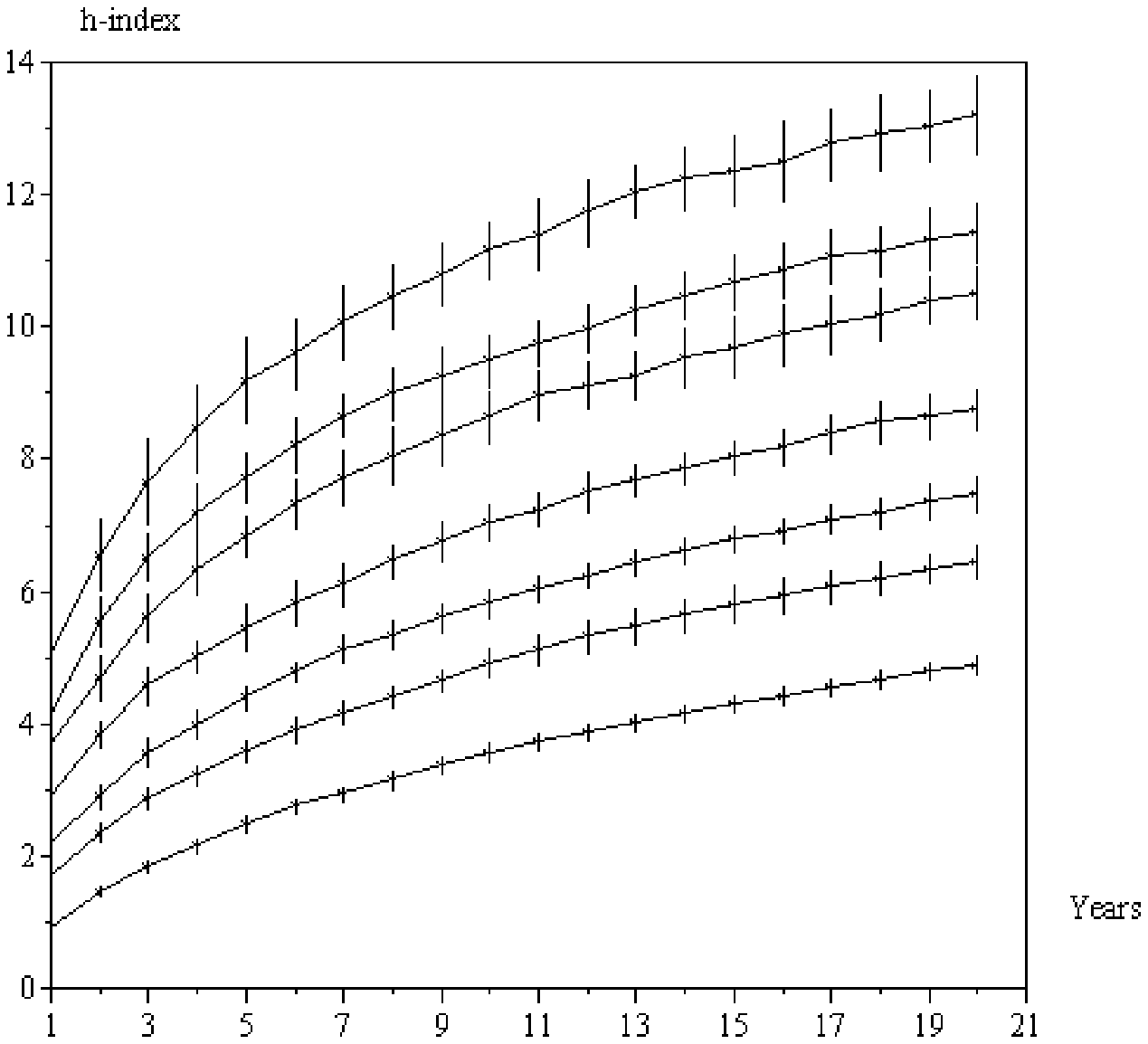}  \hspace{1cm}
   \includegraphics[scale=0.5,angle=0]{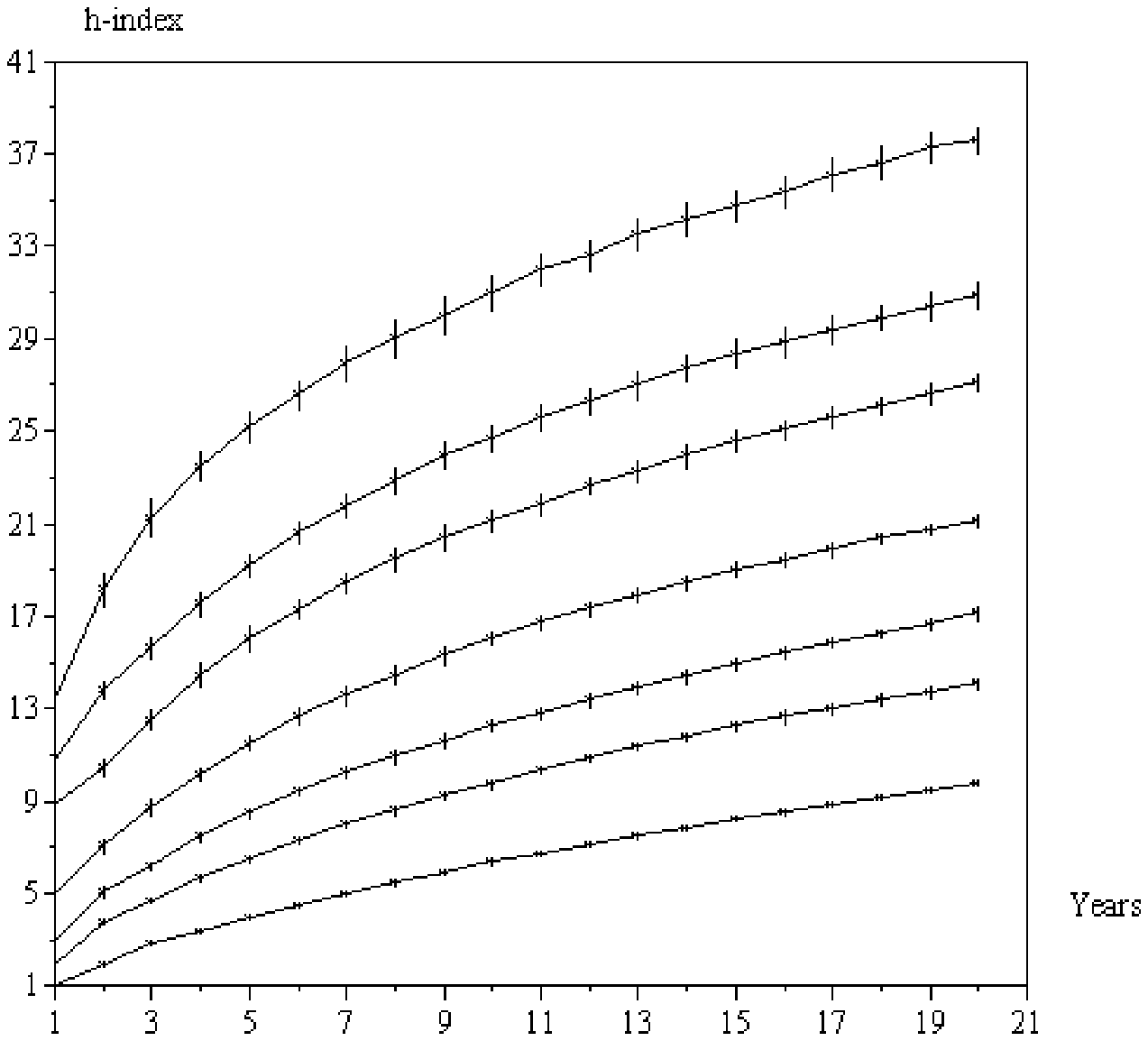} \\
   (e)   \hspace{8cm}   (f)  \\
   \caption{The $h-$indices for the seven considered types of
   authors obtained for any of the two communities with the
   \emph{UNI} model (a) and the indices obtained for communities A (c)
   and B (e) while considering the \emph{PREFF} (b) model for $w=5$. 
   The analogue results obtained for $w=20$ are given in (b) and
   (d,f)~\label{fig:ER}}
\end{center}
\end{figure*}

Figure~\ref{fig:ER})(a) shows the evolution of the $h-$indices for the
seven considered types of authors (i.e. those publishing $y = (1, 2,
3, 5, 10, 15, 30)$ articles per year have similar dynamics and are
averaged together) in community A or B under the \emph{UNI} dynamics
while assuming $w=5$. The analogue results obtained for the
\emph{PREFF} dynamics for communities A and B are given in
Figures~\ref{fig:ER}(c) and (e), respectively.
Figures~\ref{fig:ER}(b,d,f) give the respective results obtained for
$w=20$.  It is clear from Figure~\ref{fig:ER} that the $h-$indices of
all types of authors tend to increase monotonically with time, though
at different rates.  Actually, as revealed after some elementary
reasoning, all citations will tend to increase linearly with the
years.  This is a direct consequence of the adopted undiscriminate
citation scheme: in principle, any author will receive a fixed average
number of citations per year (equal to $w$).  Therefore, the
$h-$indices will be roughly proportional to the square root of the
years. In addition, the $h-$indices of each type of author will
directly reflect its yearly production.

Because of the linear rate of increase of the citations per type of
author, this model has little interest, except for providing a
comparison standard for the other models considering citations
preferential to the number of citations.  In particular, note that in
the case of identical community fitness values (shown in (a) for $w=5$
and (d) for $w=20$), the evolution of the $h-$indices would not be too
different from those obtained for different fitness values (shown in
(b-c) for $w=5$ and (e-f) for $w=20$).  For instance, the most
productive author in community A would reach an $h-$index of 13 after
20 years in case the two communities were identical and an $h-$index
of 18 after that same period in case its community had twice as much
fitness as community B.  In other words, the different fitness values
have relatively little effect on the relative evolution of the
$h-$indices.

\begin{figure*}[h]
 \begin{center}
   \includegraphics[scale=0.5,angle=0]{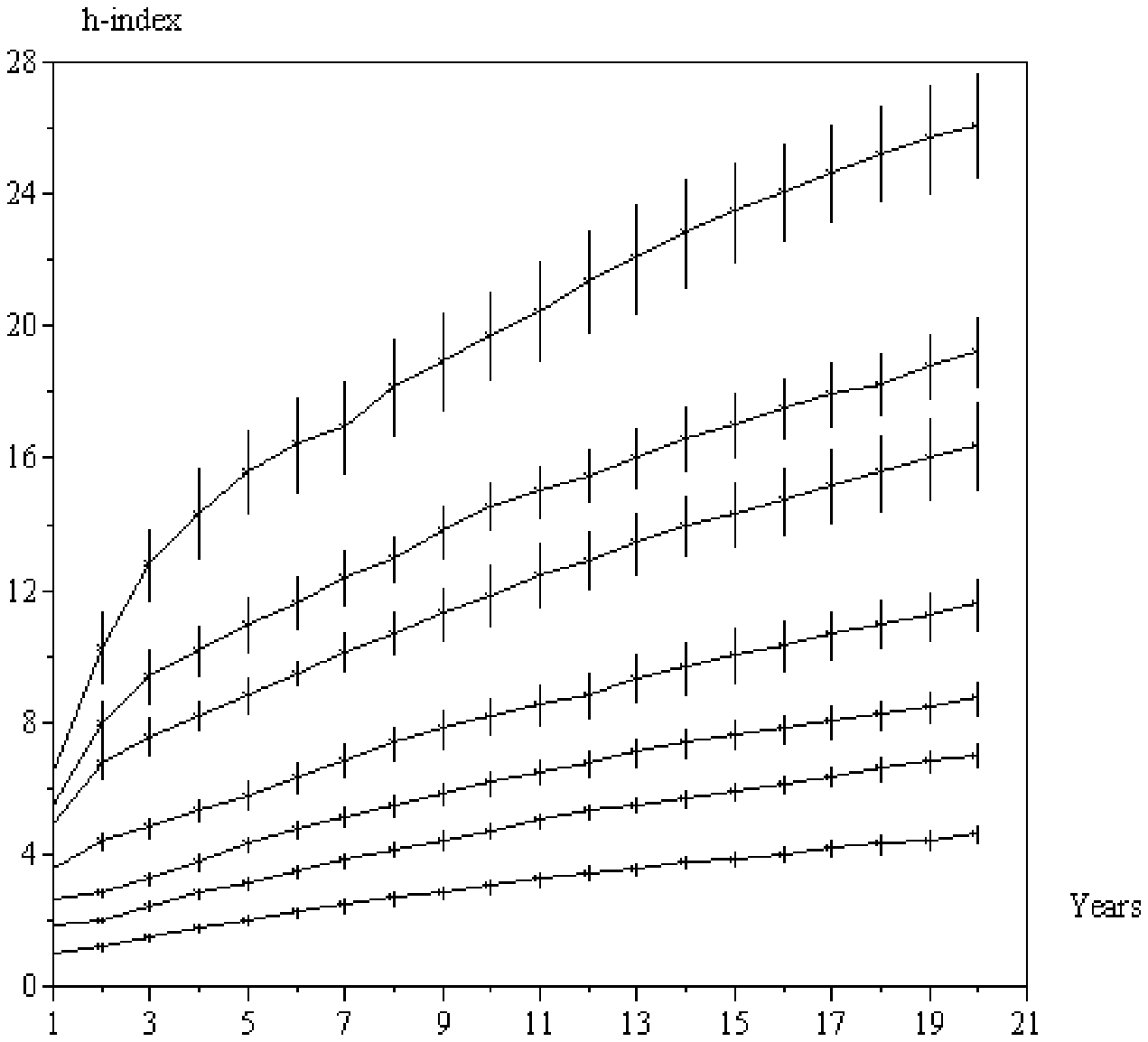} \hspace{1cm}
   \includegraphics[scale=0.5,angle=0]{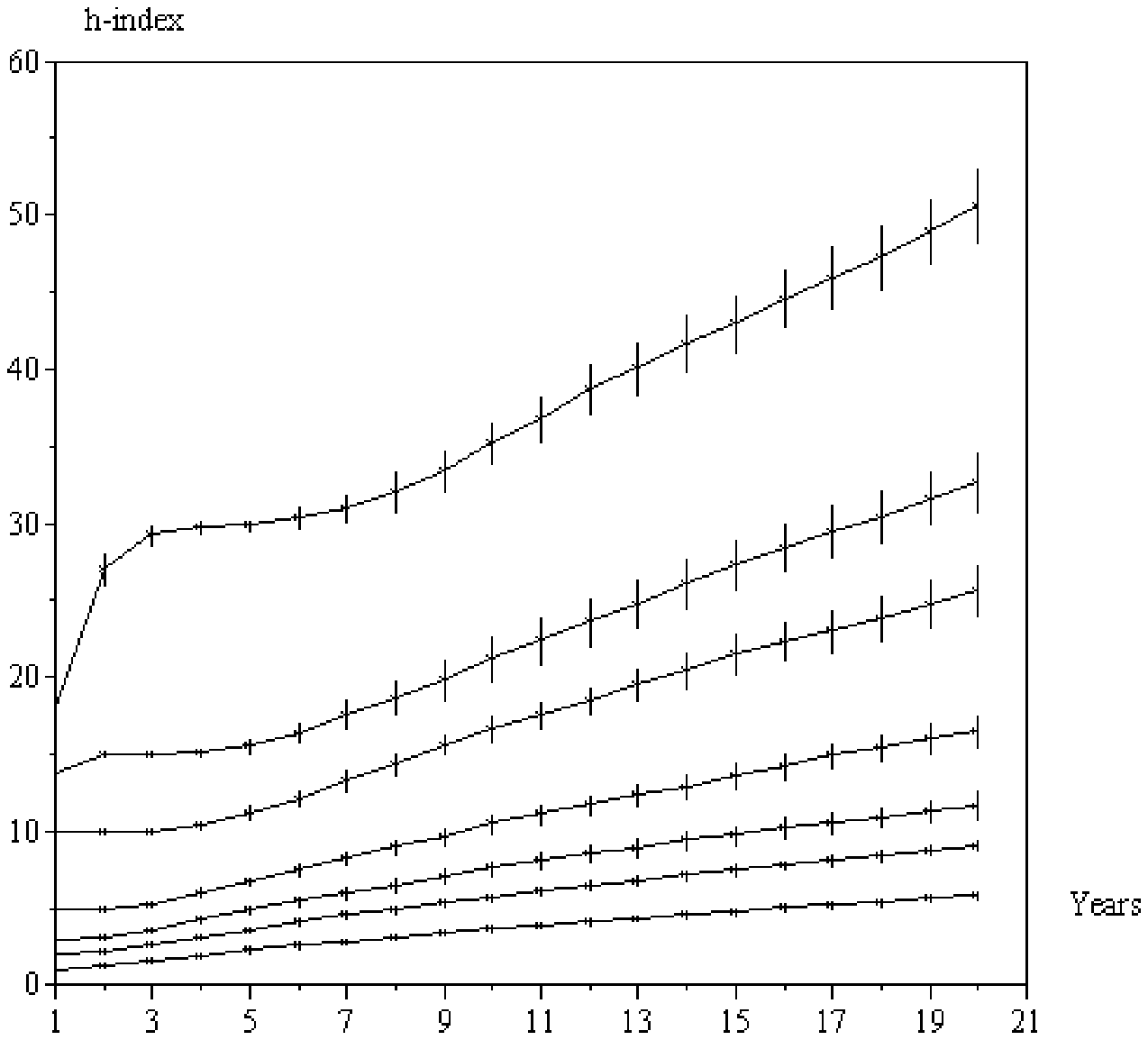} \\
   (a)  \hspace{8cm}   (b)  \\
   \includegraphics[scale=0.5,angle=0]{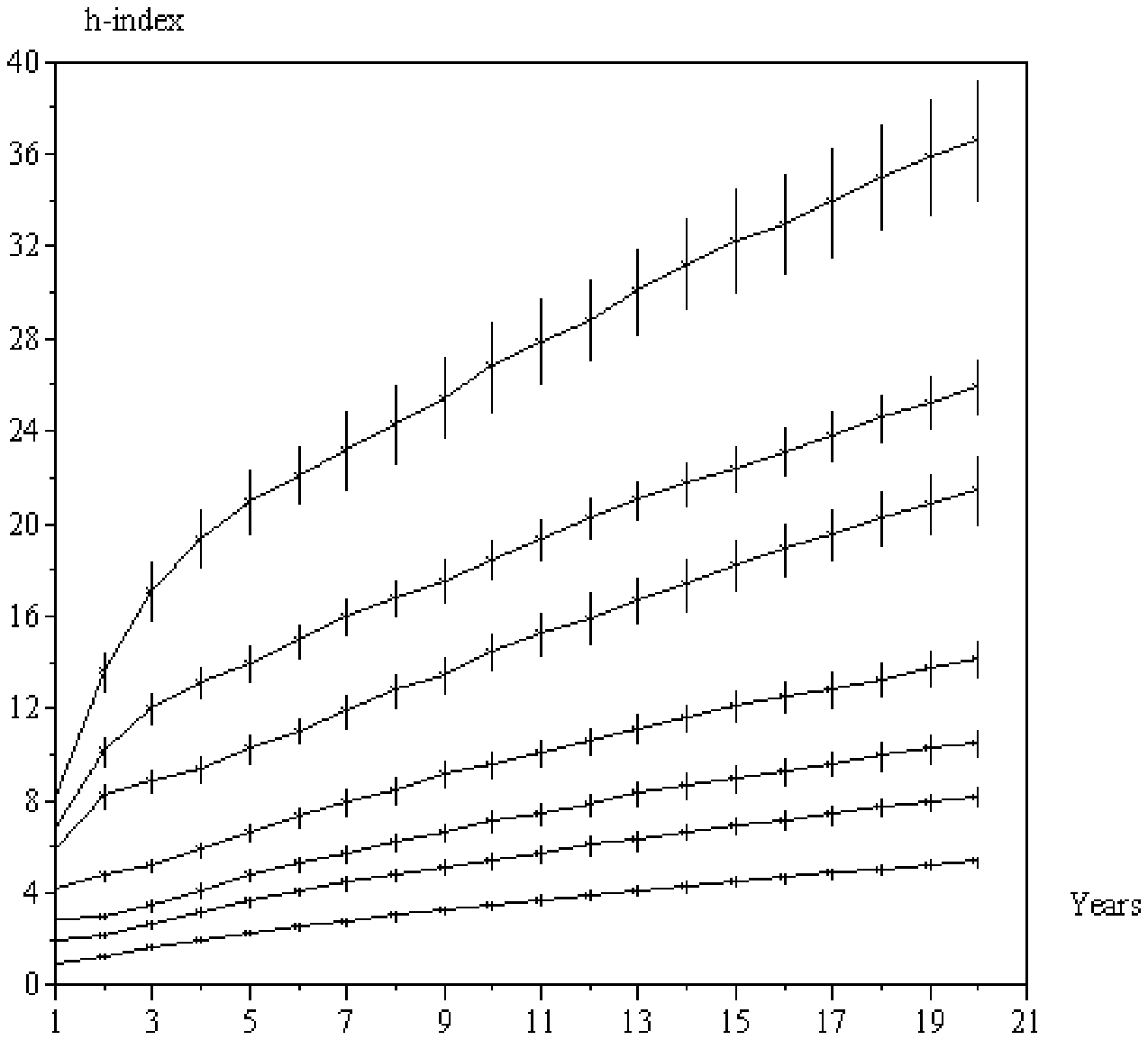} \hspace{1cm}
   \includegraphics[scale=0.5,angle=0]{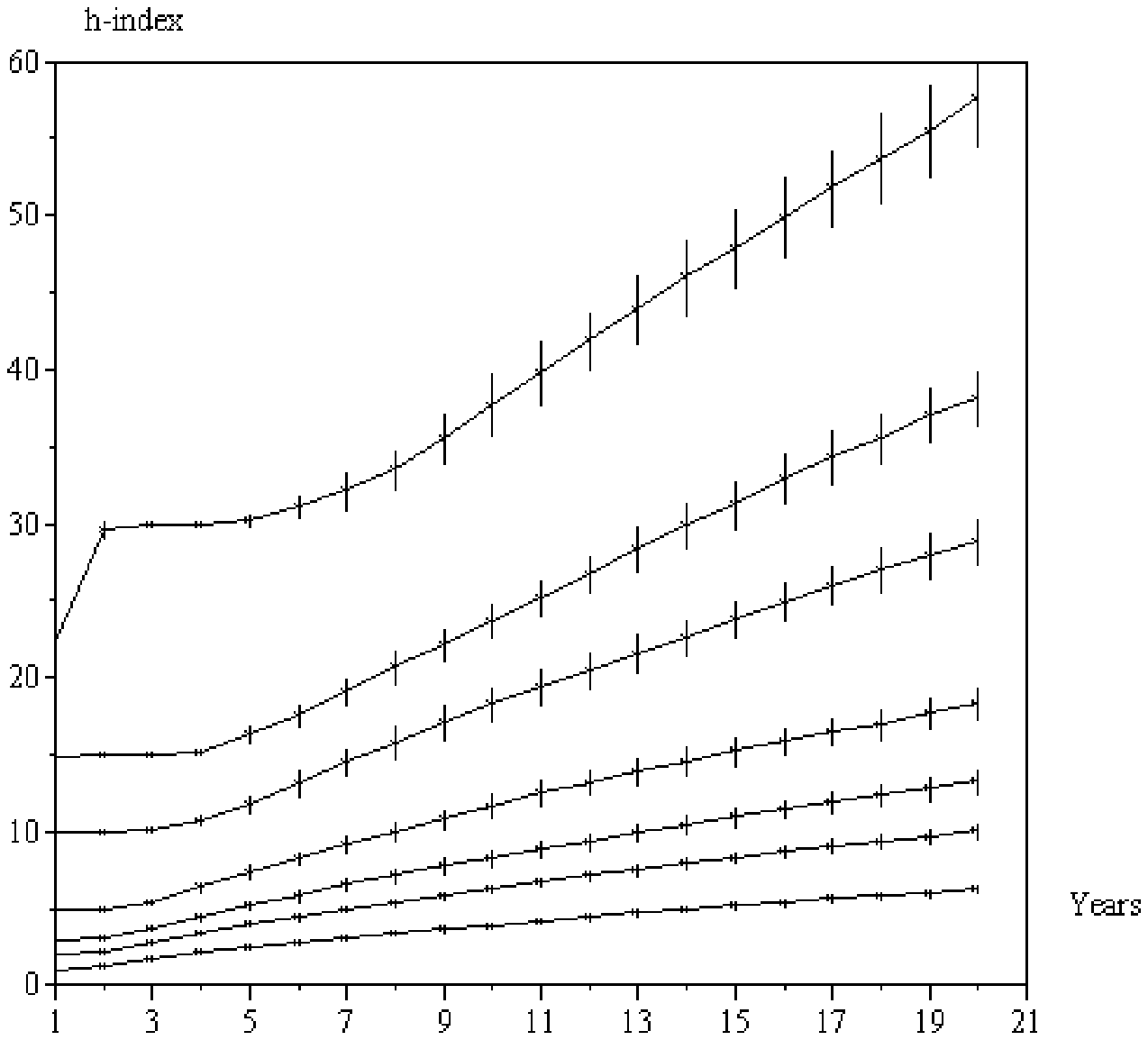} \\
   (c)  \hspace{8cm}   (d)  \\
   \includegraphics[scale=0.5,angle=0]{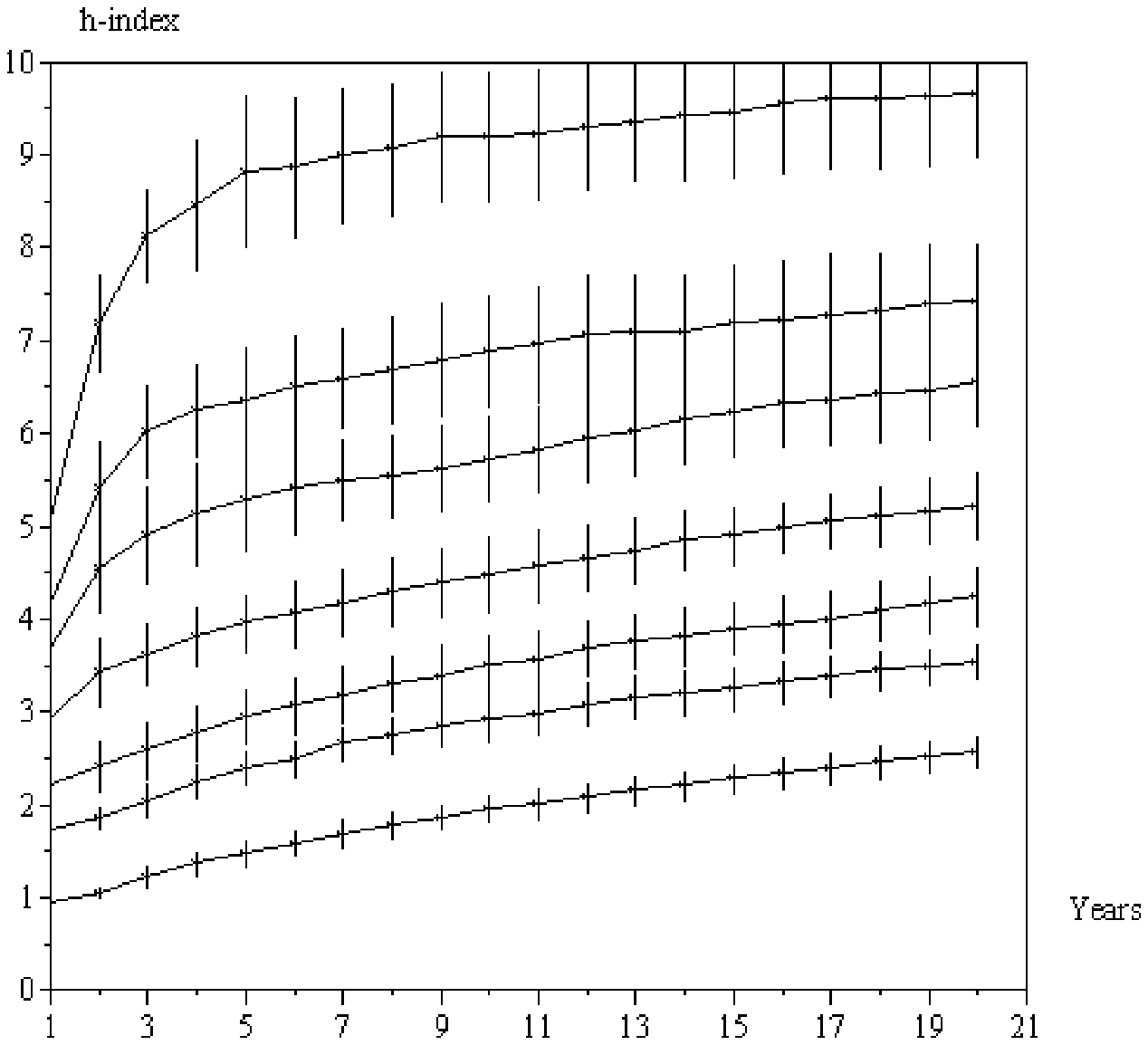} \hspace{1cm}
   \includegraphics[scale=0.5,angle=0]{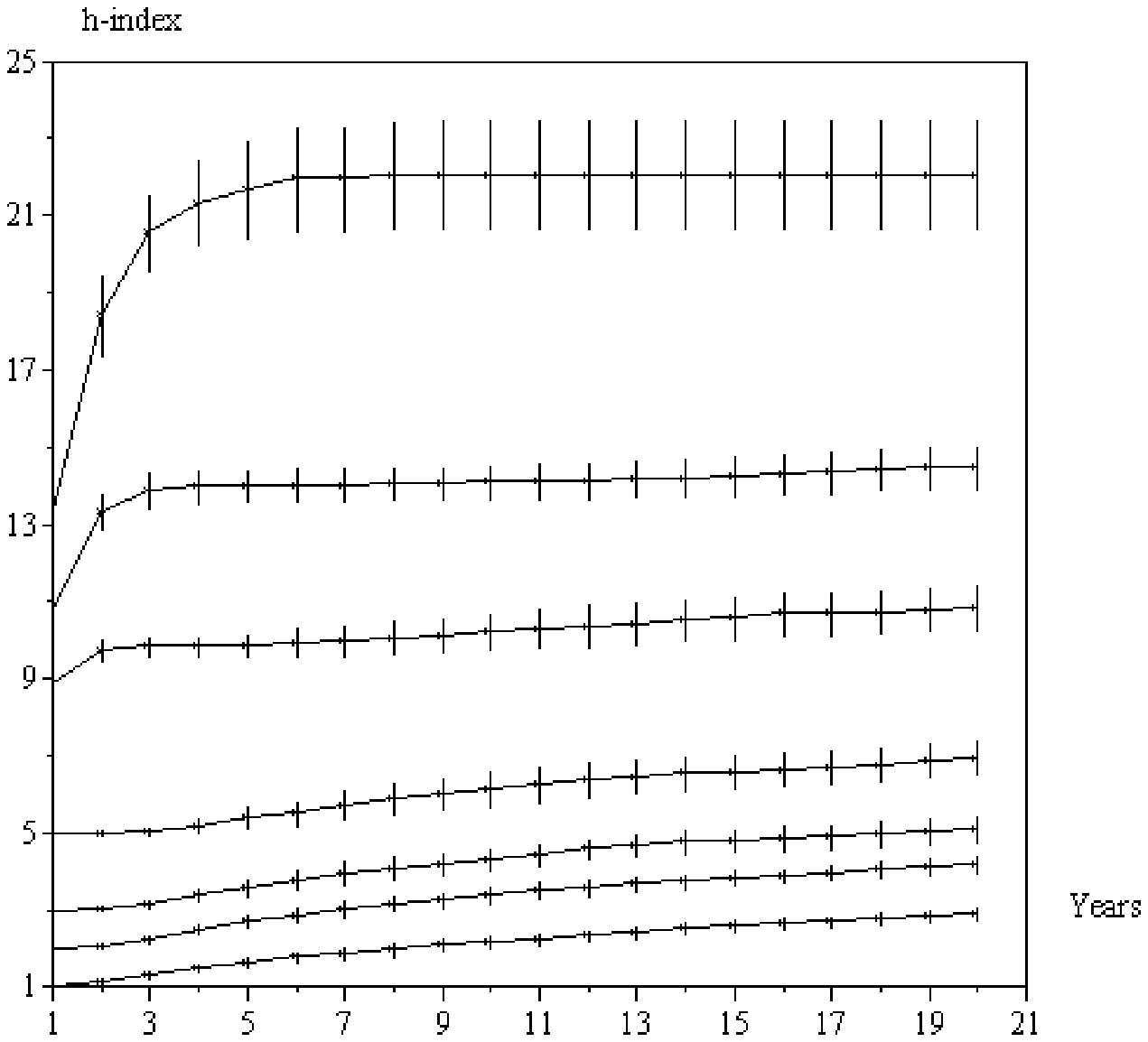} \\
   (e)  \hspace{8cm}   (f)  \\
   \caption{The $h-$indices for the seven considered types of
   authors obtained for any of the two communities with the
   \emph{PREFC} model (a) and the indices obtained for communities A (c)
   and B (e) while considering the \emph{DBPREF} (b) model for $w=5$.
   The analogue results obtained for $w=20$ are shown in (b) and
   (d,f).~\label{fig:BA}}
\end{center}
\end{figure*}

Figure~\ref{fig:BA})(a) shows the evolution of the $h-$indices for the
seven considered types of authors in communities A or B under the
\emph{PREFC} dynamics while assuming $w=5$. The analogue results obtained for
the \emph{DBPREF} dynamics for the A and B are given in
Figures~\ref{fig:BA}(c) and (e), respectively.
Figures~\ref{fig:ER}(b,d,f) give the respective results obtained for
$w=20$.  Recall that all these simulations consider citations
preferential to the current total among citations of each article
(`rich get richer'). All curves are characterized by a non-linear
portion along the first years, followed by nearly linear evolution.
Also, as in the indiscriminate case, the $h-$indices of the 7 types of
authors tend to reflect their yearly production. As could be expected,
the standard deviations for all cases tend to increase with the author
type productivity.

Let us first discuss the situation arising for $w=5$.  Note that a
pronouncedly sharper increase of the $h-$indices is verified along the
first 4 or 5 years for the most productive author types for this value
of $w$. When no distinction is made between the fitness values of the
two communities (i.e. model \emph{PREFF}) -- see Figure~\ref{fig:BA}(a), the
$h-$indices of the 7 types of authors tend to evolve steadily until
reaching, at year 20, the configuration shown in line 1 of
Table~\ref{tab:h}.  Now, in the case of different fitness values for
the two communities (model \emph{DBPREF}), the evolution of the $h-$indices is
much steeper for community A (Fig.~\ref{fig:BA}a) than for community B
(Fig.~\ref{fig:BA}b).  The $h-$indices harvested after 20 years by the
7 types of authors in communities A and B in this case would be like
those given in lines 2 and 3 of Table~\ref{tab:h}, respectively.  The
ratio between the $h-$indices of communities A and B with different
fitness values and the $h-$index values in the case of equal fitness
are given in lines 4 and 5, respectively, in Table~\ref{tab:h}.

Strikingly, while the different fitness of community A contributes
to moderate increase ratios varying from 1.174 to 1.402, the effect
is catastrophic for community B, with respective ratios varying from
0.56 to 0.37.  The reason for such a dynamics is that, with the
progress of the years, the articles in community A become ever more
cited and competitive, deviating most of the citations that would be
otherwise established within community B.  This is a situation
where, though the rich do not get so much richer, the poor becomes
irreversibly poorer as the preferential effect will continue until
virtually no citation take place yearly inside community B. An even
more acute situation would have been observed in the likely case
that the fitness of community A increased with its overall growing
$h-$indices.  As is visible in Figure~\ref{fig:BA}(e), this same
effect will slightly contribute to level the $h-$index values among
the individuals in community B.

The situation for $w=20$ is largely similar to that discussed above
for $w=5$, with the following differences.  First, a short plateau of
$h-$index values appear along the first years, especially for the most
productive authors in the cases of equal fitness
(Figure~\ref{fig:BA}b) and for community A with different fitness
(Figure~\ref{fig:BA}d).  The relative increase of the $h-$indices
observed with respect to the equal fitness case (i.e. the ratios
between the lines 7 and 8 with line 6, respectively) are given in
lines 9 and 10.  Now, while minimal increase ranging from 1.052 to
1.162 is obtained for community A in the case of different fitness
values, the ratio for community B varies from 0.50 to 0.43.  In
addition, the exhaustion of the citations inside community B is now
clearly visible in the saturation of the $h-$indices in
Figure~\ref{fig:BA}(f).

\begin{table}
  \centering
    \vspace{1cm}
     \begin{tabular}{||l|r|r|r|r|r|r|r||} \hline
       line  &  AT1  &  AT2  &  AT3  &  AT4  &  AT5  &  AT6  &  AT7  \\  \hline
              1   & 4.6   &  7.0  & 8.7   &  11.5 &  16.4 & 19.2  &  26.1 \\
              2   & 5.4   &  8.2  & 10.5  &  14.2 &  21.5 & 25.9  & 36.6  \\
              3   & 2.6   &  3.5  &  4.3  &   5.3 &   6.6 &  7.4  & 9.7   \\
              4   & 1.174 &  1.171&  1.207&  1.235&  1.311&  1.349& 1.402 \\
              5   & 0.56  &  0.50 &  0.49 &   0.46&  0.40 &  0.39 &  0.37 \\ \hline
              6   & 5.8   &  9.1  & 11.7  & 16.6  & 25.6  & 32.6  & 50.7  \\
              7   & 6.1   &  9.9  & 13.3  & 18.1  & 29.5  & 38.2  & 58.9  \\
              8   & 2.9   &  4.2  &  5.2  & 6.9   & 11.0  & 14.7  & 21.8  \\
              9   & 1.052 &  1.088&  1.137&  1.090&  1.152&  1.172& 1.162 \\
             10   & 0.50  &  0.46 &  0.44 &  0.42 &  0.43 &  0.45 & 0.43  \\ \hline
     \end{tabular}
  \caption{The $h-$indices of the 7 types of authors after 20 years and
              respective ratios.  See text for explanation.
              Each of the author types $i$ is identified as $ATi$.
              }\label{tab:h}
\end{table}

\section{Strategies for Improving Individual $h-$Indices}

Given the largely unfair dynamics identified for the authors in
community B, it becomes interesting to consider by which means this
situation could be, at least partially, improved.  Of course, in case
the fitness difference were a direct consequence of the quality of the
publications in community B, the immediate answer would be that the
authors in that community should try to improve their standards or be
doomed indeed.  However, in case the differences of fitness have a
more arbitrary and biased origin, it becomes justifiable to consider
means to correct the situation.  The following three possibilities,
which are by no means exhaustive, could be considered:

{\bf A bit more attention from the richer:} Authors in community A
tries to cite those in B more frequently.  The main advantage of this
solution is that the authors in community A would just loose a little
bit, while those in B would gain a lot with respect to the even
fitness situation.  After all, citations should be based only on the
inherent quality and contribution of each work.

{\bf Collaborative strategy:} Authors in B participate as co-authors
with community A.  Although such a practice would tend to enhance
the $h-$index values in community B, such an increase would be
limited by the high resilience of the h-index with respect to such
initiatives.

{\bf A bit more attention among the poorer:} In this case, the authors
in community B would pay greater attention to the work of their
colleagues, trying to reduce the different fitness effect on the
preferential citations.  Again, this should reflect the inherent
quality and contributions of each work.

\section{Towards More Comprehensive Citation Indices}

Although creative proposals such as the $h-$index and enhanced
variations do provide interesting advantages for measuring the
significance of scientific publishing, they can still be biased by
several factors including the presence of communities with varying
citation fitness which, as shown in the previous section, can lead to
critical situations.  It would be interesting, in the light of the
obtained results, to consider some possible modifications and
enhancements to the $h-$index, as addressed in the following.

First, we have to go back to the reasons why citations exist after
all, which include mainly: (a) establish the context of the
research; (b) provide additional information about the adopted
concepts and methods; and (c) compare methodologies or results.
However, all such cases can be conveniently unified into the
following criterion:

\begin{itemize}
        \item {\bf Citations should included in order to complement
        the work in question. As such, all citations should be
        directly related to the main aspects developed in each new
        article}.
\end{itemize}

Now, it happens that the relationship between any two articles can be
automatically inferred, to some degree of accuracy, by using
artificial intelligence methods combined with the ever increasing
online access to hight quality scientific databases and
repositories. One of the simplest approaches involves counting how
many keywords are shared by any pair of articles. In order to define
the direction of the citations (actually its \emph{causality}), the
new article would be naturally linked to older entries in the
databases.  The number of implied citations would naturally vary with
the comparison methodology and adopted thresholds, but would
nevertheless provide a less arbitrary and complete means for getting
more comprehensive and less biased citations from which the respective
$h-$index could be calculated.  Actually, after some further
reflection it becomes clear that such a citation system allows a
series of additional advantages, including:

\begin{enumerate}
\item {\bf Inherently linked to bibliographical research:}  One of
the preliminary steps in every article is to perform a reasonably
complete research on existing related works, the so-called
\emph{bibliographical search}.  It would be interesting to use the
same system(s) for both bibliographical search \emph{and} automatic
citations, ensuring consistency.
\item {\bf More substantive evaluation:} Provided good journals (e.g. with
reasonable impact factor) are considered for the databases, the
quality of the cited works would be at least partially assured.
Indeed, a given article could be more likely to be read and evaluated
by referees of a good journal than by an eventually hassled author
seeking for contextual references.  After all, citations are known
sometimes to include copies from references in related previous
articles (e.g.~\cite{Borner:2004}).
\item {\bf Avoidance of personal biases:}  Because the virtual
citations would be established from databases while considering
objective keywords, no space is left for any eventual personal biases.
\item {\bf Quantification of the quality of the work:} With the advance of
more sophisticated intelligent computer systems, it will become
possible to have the automatic citation system to try to quantify
several important qualities of an article, including originality,
clarity, grammar, and even fraud detection.
\end{enumerate}

It can not be said that automatic citation can be easily accomplished
ro that it will be fully precise from the beginning, but certainly it
can provide a second, complementary, indication to be taken into
account jointly with more traditional scientometric indices.  At the
same time, the continuing advances of multivariate statistics and
artificial intelligence will contribute to achieving ever more
intelligent and versatile automatic citation and indexing systems.

\section{Concluding Remarks and Future Works}

In order any artificial process can be improved, it is imperative to
quantify its performance in the most objective and unbiased way as
possible.  Scientific citations -- properly normalized by area, number
of authors and always under the auspices of common sense -- are no
exception to this rule.  Since the first printed scientific and
technical works, authors and readers have been involved in an ever
evolving complex system of citations aimed at contextualizing and
complementing each piece of reported research.  Though indicators such
as the total number of published articles per author, the total number
of citations, or the citations per article, amongst many others, have
been systematically used for promotions, grants and identification of
scientific trends, there is still no perfect index.  Recently
introduced by Hirsch~\cite{Hirsch:2005}, the $h-$index presents a
series of interesting advantages over more traditional indicators, as
well as some specific shortcomings which have been progressively
addressed.

At the same time as scientometrics progresses healthly and inexorably,
it is important to stick to the original aims of scientific
publication, namely the dissemination of new findings in order to
foster even further development.  In order to complement and enhance
reported works, it is essential to provide significant and unbiased
citations which can properly contextualize and complement each piece
of work.  Primarily, each citation is an acknowledgement of a previous
work, contributing to its significance and recognition of the
respective author. However, because scientometrics increasingly
determines the course of science, it is critically important to always
revise and improve the respective indices.

The present work has addressed the dynamical evolution of the
$h-$index considering a limited period of time (20 years) in a
citation network involving two communities whose number of authors
follow a particular configuration of Zipf's law.  Other distinguishing
features of the reported models include the consideration of citations
preferential to an inherent value of fitness assigned to each
community as well as to the existing number of citations.  Although
the number of papers published by year by each author remains
constant, two different number of citations emanating from each
article (i.e. $w=5$ and 20) were considered separately.

Four types of models were considered in simulations involving 50
realizations of each configuration.  Linear increase of citations was
observed for the two models involving indiscriminate citations and
citations preferential to the community fitness only.  The two more
realistic situations assuming the citations to be preferential to the
current number of citations of each paper, especially the model where
the citations were also preferential to the community fitness values,
yielded particularly interestin results.  When compared to the
evolution of the $h-$indices of the two communities evolving with
citations preferential only to the number of citations, the model
involving citations also preferential to the communities fitness
values showed that the authors in community A experienced moderate
increase in the $h-$indices while the indices of the authors in
community B suffered severe reduction.  It should be recalled that the
presence of coexisting communities is but a hypothesis, to be
eventually confirmed through additional experimental work.

Having identified such trends in multiple-community systems of
citations, we briefly discussed three strategies which could be
adopted in order to compensate for the different fitness values.  In
addition, an improved approach has been outlined which can provide
complementary characterization of the significance and productivity of
the production of authors or groups.  More specifically, it has been
suggested that statistical andartificial intelligence methods be used
in order to identify virtual citations from each new work to other
previous works stores in databases while taking into account the
overlap of key features (e.g. key words, main contributions, etc.)
between the new and previous works.  A number of further advantages
have been identified for this approach.

Future extensions of the present work include the consideration of
larger number of authors, coexistence of more than two communities, as
well as the investigation of possible border effects implied by the
relatively small size of the adopted networks.  It would also be
interesting to perform simulations taking into account longer periods
of time, citation time windows (e.g. no citations to articles older
than a given threshold), and the progressive addition and retirement
of authors.

Scientometrics corresponds to a peculiarly interesting circular
applicaton of science to improve itself through the proposal of ever
more accurate and unbiased indices and measurements. While the
advances of computing have implied an inexorably increasing number of
articles and new results, it is suggested that they also hold the key
-- in the form of artificial intelligence -- to proper quantification
of scientific productivity and quality.  After all, as hinted in the
quotation at the beginning of this work, if human attention is
becoming so scarce, perhaps automated digital attention can at least
provide some complementation.

\vspace{1cm}

Luciano da F. Costa is grateful to CNPq (308231/03-1) and FAPESP
(05/00587-5) for financial support.

\bibliographystyle{apsrev}
\bibliography{h}

\begin{thebibliography}{18}
\expandafter\ifx\csname natexlab\endcsname\relax\def\natexlab#1{#1}\fi
\expandafter\ifx\csname bibnamefont\endcsname\relax
  \def\bibnamefont#1{#1}\fi
\expandafter\ifx\csname bibfnamefont\endcsname\relax
  \def\bibfnamefont#1{#1}\fi
\expandafter\ifx\csname citenamefont\endcsname\relax
  \def\citenamefont#1{#1}\fi
\expandafter\ifx\csname url\endcsname\relax
  \def\url#1{\texttt{#1}}\fi
\expandafter\ifx\csname urlprefix\endcsname\relax\def\urlprefix{URL }\fi
\providecommand{\bibinfo}[2]{#2}
\providecommand{\eprint}[2][]{\url{#2}}

\bibitem[{\citenamefont{Garfield}(1972)}]{Garfield:72}
\bibinfo{author}{\bibfnamefont{E.}~\bibnamefont{Garfield}},
  \bibinfo{journal}{Science} \textbf{\bibinfo{volume}{178}}
  (\bibinfo{year}{1972}).

\bibitem[{\citenamefont{Ball}(2005)}]{nature_news:2005}
\bibinfo{author}{\bibfnamefont{P.}~\bibnamefont{Ball}},
  \bibinfo{journal}{Nature} \textbf{\bibinfo{volume}{436}}
  (\bibinfo{year}{2005}).

\bibitem[{\citenamefont{Batista et~al.}(2007)\citenamefont{Batista, Campitely,
  Kinouch, and Martinez}}]{Batista_etal:2007}
\bibinfo{author}{\bibfnamefont{P.~A.} \bibnamefont{Batista}},
  \bibinfo{author}{\bibfnamefont{M.~G.} \bibnamefont{Campitely}},
  \bibinfo{author}{\bibfnamefont{O.}~\bibnamefont{Kinouch}}, \bibnamefont{and}
  \bibinfo{author}{\bibfnamefont{A.~S.} \bibnamefont{Martinez}}
  (\bibinfo{year}{2007}), \bibinfo{note}{arXiv:physics/0509048}.

\bibitem[{\citenamefont{Bornmann and Daniel}(2005)}]{Bornmann:2005}
\bibinfo{author}{\bibfnamefont{L.}~\bibnamefont{Bornmann}} \bibnamefont{and}
  \bibinfo{author}{\bibfnamefont{H.-D.} \bibnamefont{Daniel}},
  \bibinfo{journal}{Scientometrics} \textbf{\bibinfo{volume}{5}}
  (\bibinfo{year}{2005}).

\bibitem[{\citenamefont{Sidiropoulos et~al.}(2005)\citenamefont{Sidiropoulos,
  Katsaros, and Manolopoulos}}]{Sidiropoulos:2006}
\bibinfo{author}{\bibfnamefont{A.}~\bibnamefont{Sidiropoulos}},
  \bibinfo{author}{\bibfnamefont{D.}~\bibnamefont{Katsaros}}, \bibnamefont{and}
  \bibinfo{author}{\bibfnamefont{Y.}~\bibnamefont{Manolopoulos}}
  (\bibinfo{year}{2005}), \bibinfo{note}{arXiv:cs;DL/0607066}.

\bibitem[{\citenamefont{Egghe}(2006{\natexlab{a}})}]{Egghe_dynamic:2006}
\bibinfo{author}{\bibfnamefont{L.}~\bibnamefont{Egghe}}
  (\bibinfo{year}{2006}{\natexlab{a}}), \bibinfo{note}{to appear}.

\bibitem[{\citenamefont{Egghe}(2006{\natexlab{b}})}]{Egghe_improvement:2006}
\bibinfo{author}{\bibfnamefont{L.}~\bibnamefont{Egghe}}, \bibinfo{journal}{ISSI
  Newsletter} \textbf{\bibinfo{volume}{2}}
  (\bibinfo{year}{2006}{\natexlab{b}}).

\bibitem[{\citenamefont{Boerner et~al.}(2004)\citenamefont{Boerner, Maru, and
  Goldstone}}]{Borner:2004}
\bibinfo{author}{\bibfnamefont{K.}~\bibnamefont{Boerner}},
  \bibinfo{author}{\bibfnamefont{J.~T.} \bibnamefont{Maru}}, \bibnamefont{and}
  \bibinfo{author}{\bibfnamefont{R.~L.} \bibnamefont{Goldstone}},
  \bibinfo{journal}{Proc. Natl. Acad. Sci.} \textbf{\bibinfo{volume}{101}}
  (\bibinfo{year}{2004}).

\bibitem[{\citenamefont{Hirsch}(2005)}]{Hirsch:2005}
\bibinfo{author}{\bibfnamefont{J.~E.} \bibnamefont{Hirsch}},
  \bibinfo{journal}{Proc. Nat. Acad. Sci.} \textbf{\bibinfo{volume}{102}}
  (\bibinfo{year}{2005}), \bibinfo{note}{arXiv:physics/0508025}.

\bibitem[{\citenamefont{Popov}(2005)}]{Popov:2005}
\bibinfo{author}{\bibfnamefont{S.~B.} \bibnamefont{Popov}},
  \bibinfo{journal}{Proc. Nat. Acad. Sci.}  (\bibinfo{year}{2005}),
  \bibinfo{note}{physics:0508113}.

\bibitem[{\citenamefont{Miller}(2006)}]{Miller:2006}
\bibinfo{author}{\bibfnamefont{C.~W.} \bibnamefont{Miller}}
  (\bibinfo{year}{2006}), \bibinfo{note}{cond-mat/0608183}.

\bibitem[{\citenamefont{Cronin and Meho}(2006)}]{Cronin_Meho:2006}
\bibinfo{author}{\bibfnamefont{B.}~\bibnamefont{Cronin}} \bibnamefont{and}
  \bibinfo{author}{\bibfnamefont{L.}~\bibnamefont{Meho}}, \bibinfo{journal}{J.
  Am. Soc. Inform. Sci. Techn.} \textbf{\bibinfo{volume}{57}}
  (\bibinfo{year}{2006}).

\bibitem[{\citenamefont{van Raan}(2005)}]{Raan:2005}
\bibinfo{author}{\bibfnamefont{A.~F.~J.} \bibnamefont{van Raan}}
  (\bibinfo{year}{2005}), \bibinfo{note}{physics/0511206}.

\bibitem[{\citenamefont{Albert and Barab\'asi}(2002)}]{Albert_Barab:2002}
\bibinfo{author}{\bibfnamefont{R.}~\bibnamefont{Albert}} \bibnamefont{and}
  \bibinfo{author}{\bibfnamefont{A.~L.} \bibnamefont{Barab\'asi}},
  \bibinfo{journal}{Rev. Mod. Phys.} \textbf{\bibinfo{volume}{74}},
  \bibinfo{pages}{47} (\bibinfo{year}{2002}).

\bibitem[{\citenamefont{Boccaletti et~al.}(2006)\citenamefont{Boccaletti,
  Latora, Moreno, Chavez, and Hwang}}]{Boccaletti:2006}
\bibinfo{author}{\bibfnamefont{S.}~\bibnamefont{Boccaletti}},
  \bibinfo{author}{\bibfnamefont{V.}~\bibnamefont{Latora}},
  \bibinfo{author}{\bibfnamefont{Y.}~\bibnamefont{Moreno}},
  \bibinfo{author}{\bibfnamefont{M.}~\bibnamefont{Chavez}}, \bibnamefont{and}
  \bibinfo{author}{\bibfnamefont{D.-U.} \bibnamefont{Hwang}},
  \bibinfo{journal}{Physics Reports} \textbf{\bibinfo{volume}{424}},
  \bibinfo{pages}{175} (\bibinfo{year}{2006}),
  \bibinfo{note}{cond-mat/0303516}.

\bibitem[{\citenamefont{Newman}(2003)}]{Newman:2003}
\bibinfo{author}{\bibfnamefont{M.~E.~J.} \bibnamefont{Newman}},
  \bibinfo{journal}{SIAM Review} \textbf{\bibinfo{volume}{45}},
  \bibinfo{pages}{167} (\bibinfo{year}{2003}),
  \bibinfo{note}{cond-mat/0303516}.

\bibitem[{\citenamefont{da~F.~Costa et~al.}(2006)\citenamefont{da~F.~Costa,
  Rodrigues, Travieso, and Boas}}]{Costa_surv:2006}
\bibinfo{author}{\bibfnamefont{L.}~\bibnamefont{da~F.~Costa}},
  \bibinfo{author}{\bibfnamefont{F.~A.} \bibnamefont{Rodrigues}},
  \bibinfo{author}{\bibfnamefont{G.}~\bibnamefont{Travieso}}, \bibnamefont{and}
  \bibinfo{author}{\bibfnamefont{P.~R.~V.} \bibnamefont{Boas}}
  (\bibinfo{year}{2006}), \bibinfo{note}{cond-mat/0505185}.

\bibitem[{\citenamefont{Newman}(2004)}]{Newman_Zipf:2004}
\bibinfo{author}{\bibfnamefont{M.~E.~J.} \bibnamefont{Newman}}
  (\bibinfo{year}{2004}), \bibinfo{note}{cond-mat/041004}.

\end{thebibliography}
\end{document}